\documentclass[onecolumn,prd,aps,amsmath,amssymb,epsfig]{revtex4}

\usepackage{graphicx}
\usepackage{epsfig}
\usepackage{subfigure}
\usepackage{color}
\usepackage{dcolumn}
\usepackage{bm}

\begin{document}
\title{Degrees of freedom and the phase transitions of two flavor QCD}
 \author{Topi {\sc K\"ah\"ar\"a}}
 \email{topi.kahara@phys.jyu.fi}
 \author{Kimmo {\sc Tuominen}}\email{kimmo.tuominen@phys.jyu.fi}
 \affiliation{Department of Physics, University of Jyv\"askyl\"a, Finland \\
Helsinki Institute of Physics, University of Helsinki, Finland}

\begin{abstract}
We study two effective models for QCD, the Nambu-Jona-Lasinio -model and the linear sigma 
model extended by including a Polyakov loop potential, which is fitted to reproduce pure 
gauge theory thermodynamics, and a coupling between the chiral fields and the Polyakov
loop. Thus the resulting models have as relevant degrees of freedom the Polyakov loop and 
chiral fields. By comparing the extended models with the bare chiral models we can conclude 
that the addition of the Polyakov loop is necessary in order to obtain both qualitatively 
and quantitatively correct results at finite temperatures. These results are extended to 
finite net quark densities, several thermodynamical quantites are investigated in detail 
and possible applications and consequences for relativistic heavy ion collision 
phenomenology are discussed.
\end{abstract}

\maketitle

\section{introduction}

Phase transitions in strongly interacting matter have been a subject of intense 
theoretical, computational and experimental research over the past decades. Apart from the 
physical values of the QCD parameters, a lot of effort has been devoted to understand 
different limits of the theory as the mass parameters values, numbers of colors and flavors 
are varied. Several qualitative and in some cases quantitative aspects have been revealed: 
for example, we know that in the absence of quarks the $SU(N)$ Yang-Mills theory has a 
global $Z_N$ symmetry \cite{Svetitsky:1982gs}, and there exists a gauge invariant operator 
charged under $Z_N$, the Polyakov loop, which can be identified as the order parameter of 
the theory. Hence, the deconfinement phase transition can be easily characterized using the 
universality arguments \cite{Polyakov:vu}. Numerical studies have confirmed this picture as 
it has been found that the deconfinement phase transition is second order when the number 
of colors is $N_{\rm c}=2$ \cite{Damgaard:1987wh}, first order for $N_{\rm c}=3$ 
\cite{Bacilieri:1988yq} (although weakly \cite{Kaczmarek:1999mm}), and presumably first 
order for $N_{\rm c}\geq 4$ \cite{Batrouni:1984vd}. 

Adding quarks to the theory changes the picture considerably. For light fermions in the 
fundamental and pseudoreal representations for $N_{\rm c}=3$ and $N_{\rm c}=2$, 
respectively, the corresponding $Z_3$ or $Z_2$ the center of the gauge group is not a good 
symmetry. However, the exact chiral symmetry of massless fermions is only little perturbed 
by small masses and the order parameter is the chiral condensate which characterizes the 
chiral phase transition. For $N_c$=3 and two massless quark flavors at finite temperature 
and zero net baryon density, the chiral phase transition is in the same universality class 
as the three dimensional $O(4)$ spin model \cite{Wilczek:1992sf}, becoming a smooth 
crossover as small quark masses are accounted for \cite{Scavenius:2000qd}. For $N_c=2$ the 
relevant universality class is that of $O(6)$ \cite{Holtmann:2003he}. 

In addition to finite temperature, one can also study the response of the QCD vacuum by 
considering finite net quark densities. 
At finite chemical potential one needs to take into account the pairing phenomena and 
superconducting phases result. However, a systematic approach based on the full QCD 
dynamics can be applied only in asymptotic densities \cite{Pisarski:1999bf} where 
asymptotic freedom simplifies the analysis. To determine the phases at intermediate 
densities relevant for phenomenological applications one needs to resort to effective 
models. Within the model studies taking into account the patterns of chiral symmetry 
breaking, one usually hopes to gain insight on the qualitative aspects of the phase 
diagram. Studies within different models have revealed that in cold dense matter, a first 
order phase transition to a superconducting phase characterized by nonzero diquark 
condensate takes place. Contrasting this with the finite temperature crossover transition 
at $\mu=0$ one concludes that in the $(T,\mu)$-plane there must exist a critical endpoint. 
A central paradigm for the first principle studies of the QCD equation of state 
\cite{eos_lattice} over the $(T,\mu)$-plane is therefore the existence, location and other 
properties of this critical point. The lattice determinations using different techniques 
have provided estimates for the location of the critical point. For two flavors see 
\cite{Fodor:2004nz} and \cite{Gavai:2004sd}. The existence of the critical point in three 
flavor QCD is currently debatable, \cite{Philipsen:2007rm,deForcrand:2007rq}, and in any 
case one should be careful in drawing any conclusions from $N_f=2$ results to the physical 
2+1 case . 

The perturbative calculations for cold quark matter at asymptotically large chemical 
potential or for hot quark gluon gas at high temperature cannot be directly applied to the 
phenomenologically relevant densities and temperatures. 
At finite temperature the perturbation expansion is known to converge poorly and reliable 
results can be obtained only at near-zero chemical potential well above the suitably 
defined transition temperature $T\geq 3T_c$, where the picture of weakly interacting 
dressed quark and gluon quasiparticles becomes correct. To address quantitative 
phenomenology for all temperatures at zero net quark density, a numerical "recipe" for 
interpolating smoothly between thermodynamics described by perturbation theory at high 
temperatures and by a resonance gas at low temperatures was proposed  in 
\cite{Laine:2006cp}, taking also into account the contribution of quarks with finite 
masses. This approach provides one with a working definition of $p(T,\{m_i\}),~i=1,\dots, 
N_f$ and of the resulting thermodynamics, but it does not yield insight to the nature of 
the underlying effective degrees of freedom and mechanisms responsible for the dynamics 
between the thermally dressed hot quarks and gluons and low temperature hadronic 
resonances. 

In this work we concentrate on the case of QCD with two light quark flavors and study a 
particular effective model description of it. This effective model is based on the 
following picture: As a function of the quark mass there exist two separate dynamical 
sectors. For massless quarks there is spontaneous breaking of exact chiral symmetry, while 
if the quarks are very heavy, and decouple from the dynamics, the center symmetry and its 
spontaneous breaking becomes relevant. In real QCD neither of these two is exact. Lattice 
calculations indicate that for quarks with finite masses, the transition is a smooth 
crossover as is also expected on the basis of universality arguments, and the transition 
can be located by measuring the expectation value of the chiral condensate. However, even 
if the discrete symmetry associated with deconfinement is broken by the presence of light 
quarks, one can still study the temperature dependence of the Polyakov loop on the lattice. 
This has been done, and one observes the Polyakov loop to rise from zero to one as 
temperature is increased from low to high values. Due to this behavior one also speaks of 
deconfining phase transition \cite{Karsch:1998qj}. Moreover, the lattice results 
\cite{Karsch:1998qj} indicate that at zero chemical potential chiral symmetry breaking and 
confinement (i.e. a decrease of the Polyakov loop) occur at the same critical temperature. 
Several attempts to explain these behaviors exist \cite{Brown:dm}. Relying only on the 
exact and approximate symmetries of the system and general effective field theory methods a 
qualitative solution to this puzzle was established in \cite{Mocsy:2003qw} based on the 
idea of transfer of information from the order parameters to non-critical fields. As a 
function of quark mass, from light to heavy quarks, this mechanism allows one to treat 
either chiral symmetry or center symmetry as the relevant one driving the transition and 
through interactions allowing also the other would-be-order parameter field to behave in a 
similar way. The framework proposed in \cite{Mocsy:2003qw} also explains the independence 
of deconfinement and chiral symmetry restoration in the case of adjoint quarks which do not 
break the center symmetry. The $(T,\mu)$ phase diagram for adjoint two color QCD was 
considered in detain in \cite{Sannino:2004ix}. 

This behavior has been reanalysed and confirmed in numerical studies of NJL model and 
linear sigma model coupled to the Polyakov loop via quarks which are integrated out in the 
random phase approximation \cite{Fukushima:2003fw,Ratti,Schaefer:2007pw}. In this work we 
study the coupling of Polyakov loop and chiral degrees of freedom described either with a 
linear sigma model or NJL model, and compare the resulting thermodynamics to that of two 
flavor QCD. Inclusion of the Polyakov loop makes the result qualitatively, and to some 
extent, even quantitatively insensitive to the underlying model used to describe the chiral 
degrees of freedom at finite temperature and small values of quark chemical potential. The 
importance of the Polyakov loop is not surprising, since the gluonic degrees of freedom are 
known to be important for the bulk thermodynamics of QCD matter when the net baryon 
densities are small. We consider also finite densities, and show that quantitative 
discrepancy between the two models increases as larger values of chemical potential are 
considered. To display this concretely, we determine the location of the critical point in 
$(T,\mu)$-plane and show how the two models lead to very different results. 

The effective models studied here, may provide input to the phenomenology of relativistic 
heavy ion collisions. Recently, with the advent of RHIC data, it has been established that 
the spacetime evolution of the hot dense QCD matter is well described by nearly ideal 
hydrodynamics. This means in particular that the system evoles along the lines of constant 
$S/N$. Hence, we study in particular the behavior of isentropic lines in the $(T,\mu)$ 
phase diagram in these two models. Whether these have the tendency to focus on the critical 
point is important for the possible experimental discovery of the critical point in heavy 
ion collisions. We find that in these models strong focusing behavior does not exist. 

We introduce the models in some detail in section \ref{section_models}, and carry out an 
analysis of the thermodynamics in section \ref{section_termo}. Comparing these models at 
zero chemical potential, we find that they imply very similar results. We also compare with 
the results of the resummed perturbation theory \cite{Laine:2006cp}. At finite chemical 
potential we find that the quantitative results of the models show large deviations.  
Especially the location of the QCD critical point cannot be estimated reliably within these 
models. We end with concluding remarks and discussion of further prospects in section 
\ref{section_checkout}.

\section{Models}
\label{section_models}

The chiral dynamics of two-flavor QCD is often formulated in terms of a linear or 
non-linear sigma model, which treats the Goldstone bosons as the relevant degrees of 
freedom. Of these two possibilities, the linear representation is more useful to study the 
finite temperatures and densities in order to find the phase diagrams of the theory, since 
also the order parameter is included explicitly. Yet another possibility is to treat the 
fundamental fermion fields as basic degrees of freedom, the mesons appearing as the bound 
states of the theory, and this leads to Nambu--Jona-Lasinio (NJL) -models. The effects of 
small quark masses are taken into account in the effective model Lagrangians by terms 
explicitly breaking chiral symmetry. These terms, appearing with small coefficients, render 
the chiral symmetry of the theory only approximate. Both of the above mentioned effective 
models for the phenomenology of two flavor QCD can be parametrized to describe equally well 
the vacuum structure at $T=\mu=0$. 

Finite temperature dynamics of $SU(N)$ pure gauge theory on the other hand is represented 
by $Z_N$ symmetric effective theory for which the order parameter is the Polykov loop. 
Polyakov loop can be constructed and studied also in a theory with quarks eventhough the 
presence of fermions in the fundamental representation of the gauge group breaks the center 
symmetry explicitly due to the antiperiodic boundary conditions of the fermion fields at 
finite temperature. 

Therefore, in real QCD neither chiral symmetry or the center symmetry is exact, and we know 
that the finite temperature phase transition at $\mu=0$ is a smooth crossover. However, one 
may ask which of the two symmetries is more accurate and would act as a "driving force" for 
the transition. Since chiral symmetry breaking is proportional to $m_q$, and $Z_N$ breaking 
is proportional to $1/m_q$, in the case of two light flavors, it seems natural to consider 
the system to have an approximate chiral symmetry. This expectation is strenghtened also by 
looking at the spectrum of the bound states, as the pions clearly show the approximate 
Goldstone behavior. Based on these motivations, in \cite{Mocsy:2003qw} the situation was 
considered taking the $m_q=0$ limit in which the chiral symmetry becomes exact, while the 
$Z_N$ symmetry is completely broken. Then the general principles of effective theory 
dictate a following form for the potential
\begin{eqnarray}
{\mathcal{L}}[\sigma,\pi^a,\phi]={\mathcal{L}}_0[\sigma,\pi^a]+{\mathcal{L}}_0[\phi]+
{\mathcal{L}}_{\rm{int}}[\sigma,\pi^a,\phi],
\end{eqnarray}
where ${\mathcal{L}}_0[\sigma,\pi^a]$ is the chiral lagrangian which has exact chiral 
symmetry, ${\mathcal{L}}_0[\phi]$ is the potential for the Polyakov loop and contains both 
$Z_N$ symmetric and symmetry violating terms and finally ${\mathcal{L}}_{\rm{int}}$ is the 
part containing the interactions between the chiral fields and the Polyakov loop. As shown 
in \cite{Mocsy:2003qw}, the most important term for the dynamics is $\phi(\sigma^2+\pi^2)$, 
which leads to transfer of information between the order parameter and a non-order 
parameter field. When quark mass is increased away from the chiral limit, the transition 
becomes a smooth crossover, but the coincidence of the chiral symemtry restoration and 
deconfinement is expected as long as chiral symmetry remains good approximation. 

On the other hand one can consider infinitely heavy quarks, i.e. the pure gauge limit. Then 
the mechanism described above works similarly, but the roles of chiral symmetry and center 
symmetry are switched and the deconfinement order parameter drives the change of the chiral 
condensate. Again the two phenomena will coincide.
Decreasing the quark mass from the pure gauge limit, the first order deconfinement line is 
expected to terminate at a critical point a some value of quark mass and for smaller values 
become a smooth crossover. Since lattice investigations find coincidence of critical 
temperatures related to chiral symmetry restoration and deconfinement for the accessible 
quark masses, it is reasonable to expect that there is a single phase border in $(m_q,T)$ 
plane interpolating between these well known small and large quark mass behaviors. The 
results of \cite{Mocsy:2003qw} can be applied to understand the behaviors near either small 
or large quark mass critical points. For intermediate values, more specific model studies 
or first principle lattice calculations are needed. 

In \cite{Fukushima:2003fw,Ratti} a specific model framework claimed applicable for all 
values of $m_q$ was proposed. 
In this work we study this framework in detail. We consider, side-by-side, both the NJL 
model and the linear sigma model (LSM) for two mass-degenerate quark flavors coupled to the 
Polyakov loop. The important feature underlying the dynamics in the approaches 
\cite{Fukushima:2003fw,Ratti,Schaefer:2007pw} 
is the assumption of independent deconfinement and chiral symmetry restoration described by 
the order parameters $\phi$ and $\sigma$, respectively, and having independent effective 
potentials, $U_\phi$ and $U_\sigma$, connected by interactions between the two. The central 
further assumption, then, is that the proposed interaction term yields the correct form for 
the resulting effective potential at all values of the quark mass, i.e. interpolates 
correctly between the limits of exact center symmetry and exact chiral symmetry hence 
describing also correctly the behaviors at the point $(m_{\rm{phys}},T)$, corresponding to 
real two-flavor QCD. We aim to test this underlying assumption in detail by cartographing 
the thermodynamics of these models over the $(T,\mu)$ plane and study in quantitative 
detail these two models against each other as well as against the numerical knowledge of 
real two-flavor QCD at zero net quark density. Let us start by describing the details of 
the models we use. To derive the grand canonical potential, we consider the following 
Lagrangian

\begin{eqnarray}
{\mathcal{L}}={\mathcal{L}}_{\rm{chiral}}+U_\phi,
\end{eqnarray}

where we have separated the contributions of chiral degrees of freedom and the Polyakov 
loop. The part ${\mathcal{L}}_{\rm{chiral}}$ is for the linear sigma model and NJL model, 
respectively,
\begin{eqnarray}
{\mathcal{L}}_{\rm{chiral}} &=& \bar{q}(i\gamma^\mu(\partial_\mu-ig_sA_0\delta_{\mu 
0})-g(\sigma+i\gamma_5\vec{\tau}\cdot\vec{\pi}))q
-\frac{\lambda^2}{4}(\sigma^2+\pi^2-v^2)^2+H\sigma, {\rm{~~LSM}} 
\label{lsm_lagrangian}\\
{\mathcal{L}}_{\rm{chiral}} &=&
\bar{q}(i\gamma^\mu(\partial_\mu-ig_sA_0\delta_{\mu 0})-m_0)q-\frac{(M-m_0)}{2G}, 
{\rm{~~NJL}}
\label{njl_lagrangian}
\end{eqnarray}
where $q=(u,d)$ is the light quark field, $m_0$ the bare quark mass ($m_0=m_u=m_d$, i.e 
exact isospinis assumed), $\sigma$ and $\vec{\pi}^T=(\pi^1,\pi^2,\pi^3)$ constitute a 
chiral field $\Sigma^T=(\sigma,\vec{\pi})$ and finally $M=m_0-G\langle\bar{q}q\rangle$. We 
work under the mean field approximation, hence the kinetic term of the chiral field is 
neglected in (\ref{lsm_lagrangian}) and the four-fermion interaction of the NJL-model 
lagrangian has been linearized in the condensate in (\ref{njl_lagrangian}). The symmetry 
breaking field $H$ in (\ref{lsm_lagrangian}) is $H=f_\pi m_\pi^2$, where $f_\pi=0.093$ GeV 
and $m_\pi=0.138$ GeV. The coupling $\lambda^2$ is determined by the tree level mass 
$m_\sigma^2=2\lambda^2f_\pi^2+m_\pi^2$, which is set to be 0.60 GeV. In vacuum the 
expectation values of the fields are $\sigma=f_\pi$ and $\pi=0$. Requiring that the 
constitutent mass in vacuum is about 1/3 of the nucleon mass yields $g=3.3$. In the NJL 
model the bare quark mass $m_0$ is taken to be 5.5 MeV and the coupling $G=10.08 $ 
GeV$^{-2}$. For a summary of the parameter values see
table \ref{parametertable}.

The Polyakov loop is included through the mean field potential
\begin{eqnarray}
U(\phi,\phi^\ast,T)/T^4=-\frac{b_2(T)}{2}|\phi|^2-\frac{b_3}{6}(\phi^3+\phi^{\ast 
3})+\frac{b_4}{4}(|\phi|^2)^2,
\label{polyakov_potential}
\end{eqnarray}
where
\begin{eqnarray}
b_2(T)=a_0+a_1\frac{T_0}{T}+a_2(\frac{T_0}{T})^2+a_3(\frac{T_0}{T})^3,
\end{eqnarray}
and the constants $a_i$,$b_i$ are fixed to reproduce pure gauge theory thermodynamics with 
phase transition at $T_0=270$ MeV. We adopt the values determined in \cite{Ratti} and shown 
for completeness in table \ref{parametertable}. Here $\phi$ is the gauge invariant Polyakov 
loop in the fundamental representation. One could also include other loop degrees of 
freedom, say, adjoint or the sextet. Here we choose to start with the mean field potential 
of the fundamental loop parametrized to describe the pure gauge thermodynamics and study 
how the interactions with the chiral degrees of freedom affect it and compare against full 
QCD thermodynamics. The extensions towards other possible degrees of freedom we leave for 
future work. Recently there has been interesting developements in the construction of 
effective theory for the pure gauge thermodynamics, e.g. 
\cite{Vuorinen:2006nz,Pisarski:2006hz}, which could be in principle coupled to chiral 
fields to obtain an effective theory for QCD. Here we choose to remain with the mean field 
potential (\ref{polyakov_potential}) which seems to capture the pure gauge thermodynamics 
sufficiently well for our purposes.

Then, for a spatially uniform system in thermodynamical equilibrium at temperature $T$ and 
quark chemical potential $\mu$ the partition function is
\begin{eqnarray}
{\mathcal{Z}} &=& {\rm{Tr}}\exp[-({\mathcal{H}}-\mu{\mathcal{N}})]\nonumber \\
 &=& \int {\mathcal{D}}\bar{q}{\mathcal{D}}q\exp[\int_x({\mathcal{L}}+\mu\bar{q}\gamma^0 
q)]. 
\end{eqnarray}
The integration over the spacetime in the action is over the compact euclidean time 
direction and over the spatial three-volume $V$. 
Since the action is quadratic in quark fields, the functional integral is easily performed 
with standard methods leading to the grand canonical potential
\begin{eqnarray}
\Omega &=& -\frac{T\ln{\mathcal{Z}}}{V} \nonumber\\
 &=& U_{\rm{chiral}}+U_\phi+\Omega_{\bar{q}q},
\end{eqnarray}
 where
\begin{eqnarray}
U_{\rm{chiral}} &=& \frac{\lambda^2}{4}(\sigma^2+\pi^2-v^2)^2-H\sigma, 
{\rm{~~LSM}}\nonumber \\
U_{\rm{chiral}} &=& \frac{(m_0-M)^2}{2G}, {\rm{~~NJL}}
\end{eqnarray}
for the chiral contribution and $U_\phi=U(\phi,\phi^\ast,T)$. The interactions between the 
Polyakov loop and chiral degrees of freedom are 
\begin{eqnarray}
\label{omegaqqbar}
\Omega_{\bar{q}q} &=& 
-2N_fT\int\frac{d^3p}{(2\pi)^3}\left({\rm{Tr}}_c\ln[1+Le^{-(E-\mu)/T}]+{\rm{Tr}}_c\ln[1
+L^\dagger 
e^{-(E+\mu)/T}]\right)-6N_f\int\frac{d^3p}{(2\pi)^3}E\theta(\Lambda^2-|\vec{p}|^2),
\end{eqnarray}
where the trace over color remains, $E=\sqrt{\vec{p}^2+M^2}$ (and further $M=g\sigma$ in 
LSM). In LSM we neglect the vacuum contribution term in $\Omega_{\bar{q}q}$ and in the NJL 
model 
it is controlled by the cutoff $\Lambda$ as indicated in (\ref{omegaqqbar}).
Performing the remaining trace gives
\begin{eqnarray}
&\left({\rm{Tr}}_c\ln[1+Le^{-(E-\mu)/T}]+{\rm{Tr}}_c\ln[1
+L^\dagger e^{-(E+\mu)/T}]\right) = \nonumber\\
 &\ln(1+3(\phi+\phi^\ast e^{-(E-\mu)/T})e^{-(E-\mu)/T}+e^{-3(E-\mu)/T})
+\ln(1+3(\phi^\ast+\phi e^{-(E+\mu)/T})e^{-(E+\mu)/T}+e^{-3(E+\mu)/T}).
\end{eqnarray}

Note that in principle the chemical potential affects the Polyakov loop potential directly, 
see \cite{Schaefer:2007pw}, but we will not be considing these effects. Having determined 
the grand canonical potential both for the Polyakov loop linear sigma model (PLSM) and 
Polykov loop NJL model (PNJL), the thermodynamics is now determined by solving the 
equations of motion for the mean fields,
\begin{eqnarray}
\frac{\partial\Omega}{\partial\sigma}=0, ~~\frac{\partial\Omega}{\partial\phi}=0, 
~~\frac{\partial\Omega}{\partial\phi^\ast}=0,
\end{eqnarray}
and then the pressure is given by evaluating the potential on the minimum: 
$p=-\Omega(T,\mu)$. We now proceed to solve numerically the thermodynamics of PLSM and PNJL 
models and compare them with each other as well as against the numerical results on 
two-flavor QCD at zero chemical potential.

\begin{table}
\begin{tabular}{c | c | c | c}
\hline
{\bf{LSM:}} & $v = f_\pi$ & $\lambda$ & $g$ \\
 & 0.093 GeV & 4.44& 3.3 \\
\hline
{\bf{NJL:}} & $m_0$ & $\Lambda$ & $G$ \\
& 5.5 MeV & 651 MeV & 10.08 (Gev)$^{-2}$ \\
\hline 
{\bf{Polyakov:}} & $a_0$ & $a_1$ & $a_2$ \\
 & 6.75 & -1.95 & 2.625\\
 & $a_3$ & $b_3$ & $b_4$ \\
 & -7.44 & 0.75 & 7.5 \\
 \hline
\end{tabular}
\caption{The parameters used for the effective potential}
\label{parametertable}
\end{table}

\section{Numerical results}
\label{section_termo}

\subsection{Thermodynamics at $\mu=0$, comparison to QCD}

\begin{figure}[htb]
\centering
  \subfigure{
  \includegraphics[width=8.5cm]{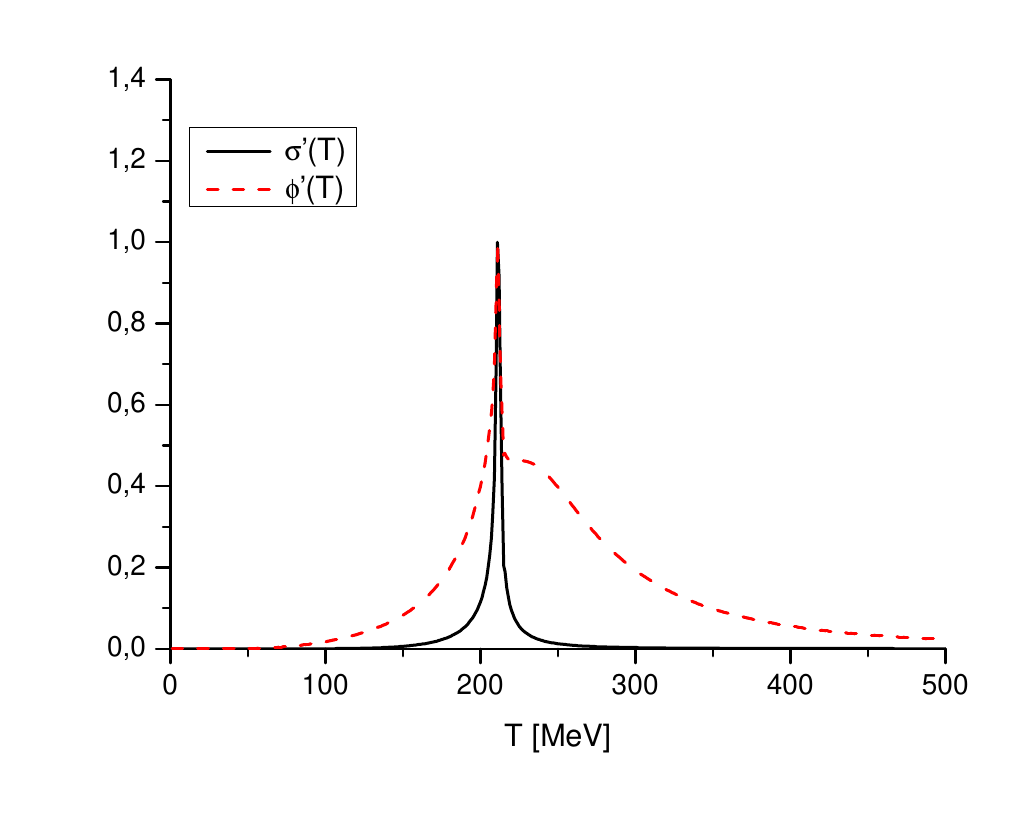}}
	\qquad
	\subfigure{\includegraphics[width=8.5cm]{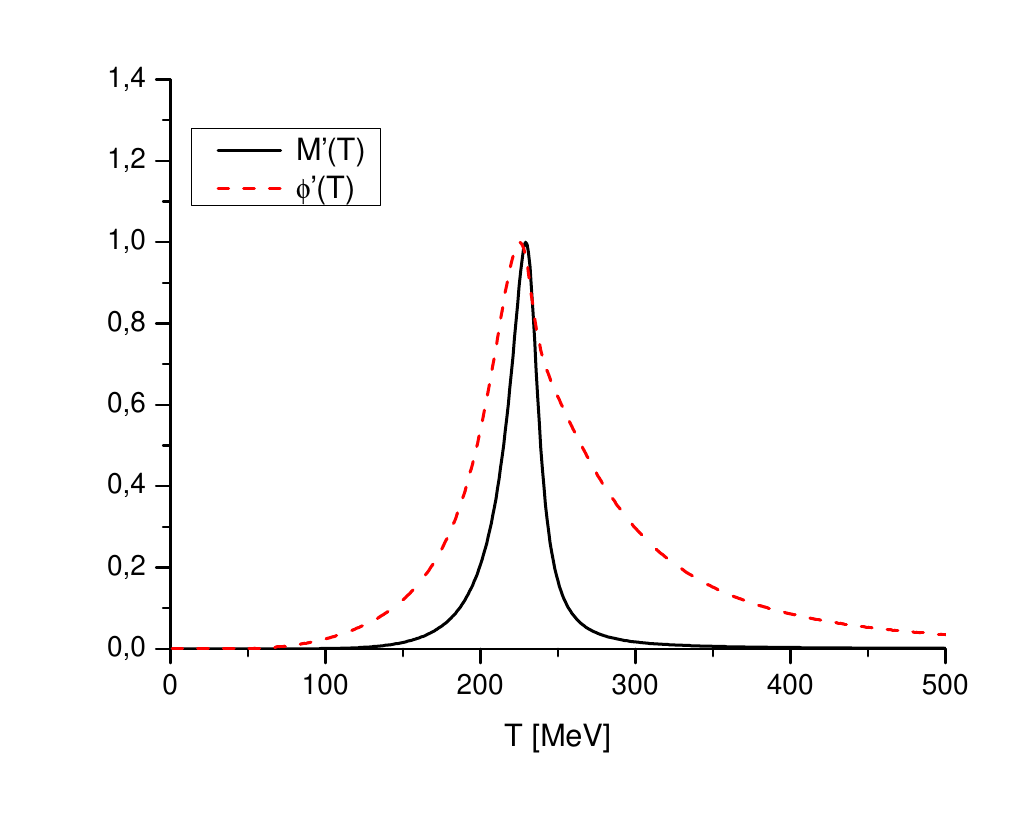}}
\vskip-0.4truecm
\caption{{\it Left Panel}: Temperature derivatives of the mean fields $\phi(T)$ and 
$\sigma(T)$ at $\mu=0$ in the PLSM model.  {\it Right Panel:} Same observables as in the 
left panel but in the PNJL model.}
\label{kondensaatit}
\end{figure}

Let us first consider the models at zero chemical potential but finite temperature. 
When finite temperature is considered, it is well known that the quantitive results of 
chiral effective theories differ: While NJL model predicts chiral restoration at $T\sim 
150$ MeV, the linear sigma model leads to result $~T\sim 190$ MeV \cite{Scavenius:2000qd}; 
allowing for finite chemical potentials only widens the spread. Including the Polyakov loop 
has important consequence as now both of these models predict a crossover near $T\sim 
210\dots 230$ MeV within 20 MeV of each other, as we show in Fig. \ref{kondensaatit}, where 
we plot the temperature derivatives of the condensates. The location of the peak defines a 
critical temperature around which the crossover takes place.  

Then consider the pressure. The coupled models, PNJL and PLSM, have already been shown to 
agree with lattice data at and above $T_c$ fairly well \cite{Ratti,Schaefer:2007pw} with 
some fine-tuning. Namely, lattice data implies a critial temperature $T_c\sim 175$ MeV, a 
value $\sim20$\% lower than we obtain. In \cite{Ratti} it has been noted that better 
agreement can be achieved in PNJL model by detuning the Polyakov loop potential away from 
the pure gauge thermodynamics through shifting of the parameter $T_0$ down to 190 MeV. 
Since the lattice data has still some uncertainty to it due to the extrapolation to 
continuum limit, we choose not to aim for perfect fits and rather plot the thermodynamical 
quantities as a function of $T/T_c$ when comparing different models. Actually, the value of 
$T_c$ should be determined by allowing for additional degrees of freedom below $T_c$ not 
considered in these effective models. We have not included the finite temperature 
contributions of the pions and more massive resonances, since we do not have a dynamical 
way to decouple their contribution at high temperatures. However, our aim is to study the 
interplay of chiral fields and Polyakov loop and the resulting thermodynamics for the 
temperature range $T_c<T<3 T_c$ for which the coupled models under consideration seem to 
work well. 

\begin{figure*}[htb]
\centering
	\subfigure{
  \includegraphics[width=8.5cm]{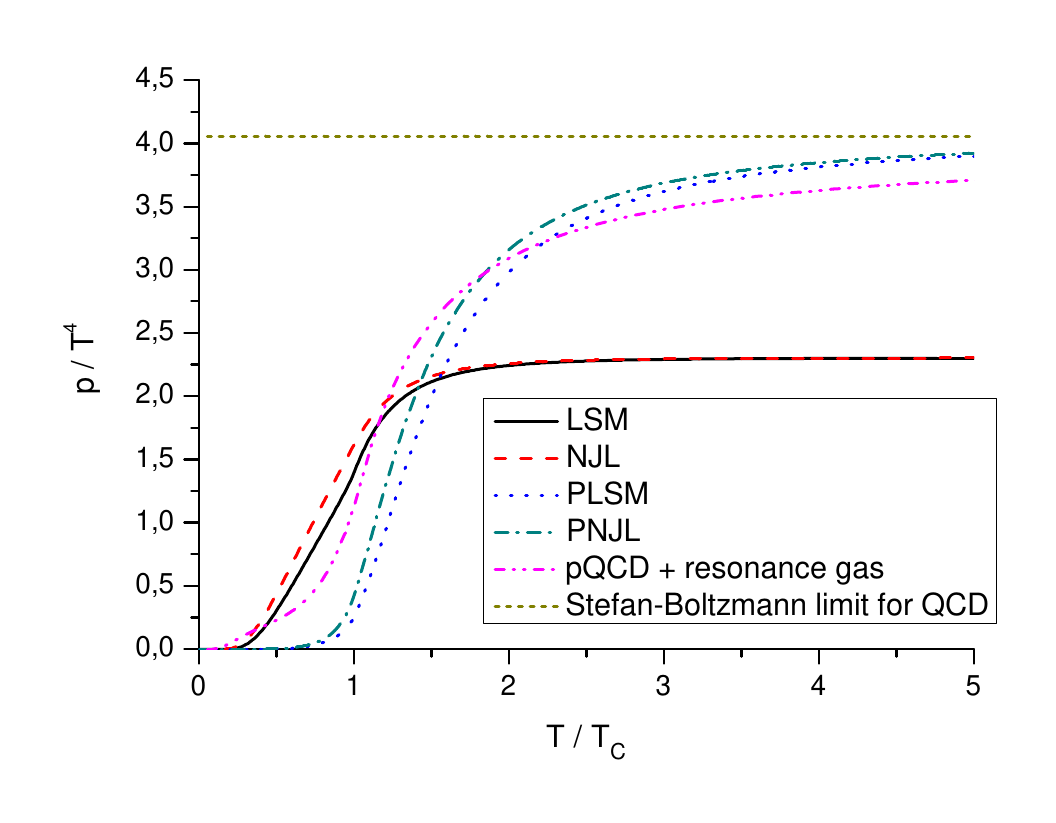}}
	\subfigure{
  \includegraphics[width=8.5cm]{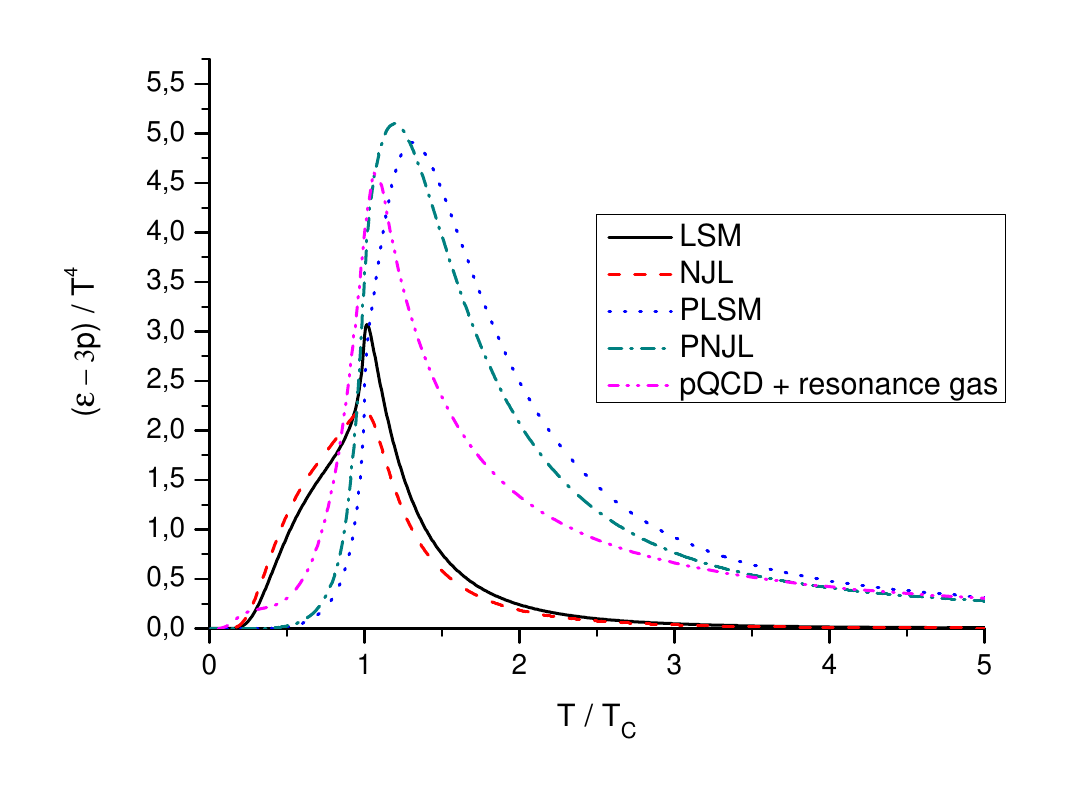}}  
\vskip-0.4truecm
\caption{{\it Left panel}: Pressure from the models with and witout the Polyakov loop at 
$\mu=0$. Also shown is the curve interpolating between the resonance gas and resummed 
perturbation theory results as well as the constant corresponding to the Stefan--Boltzmann 
limit of two-flavor QCD. {\it Right panel}: Similar figure for the trace anomaly 
$(\epsilon-3P)/T^4$.}
\label{pressure_traceanomaly}
\end{figure*}

Here, to compare with QCD, we use the numerical result of \cite{Laine:2006cp} in which the 
high temperature part of the curve is  based on the full ${\mathcal{O}}(g^2\log g)$ 
calculation in pure gauge theory, supplemented with a more phenomenological "recipe" to 
include the contribution of $N_f$ massive quarks at order ${\mathcal{O}}(g^2)$. At low 
temperatures this result is matched smoothly on the resonance gas result from 
\cite{Karsch:2003zq}. This QCD+resonance gas 
result is shown in the left panel of Fig. \ref{pressure_traceanomaly} at $N_f=2$ 
appropriate for this work \footnote{We thank M.Laine for providing the numerical results of 
\cite{Laine:2006cp} for $N_f=2$ case.}. Also shown are our results for the pressure obtained 
from the coupled models PLSM and PNJL as a function of $t=T/T_c$. For comparison we show 
the corresponding results from LSM and NJL models without the Polyakov loop. Observe how 
the inclusion of the Polyakov loop increases the result for the absolute value of the 
pressure by roughly 80 \%. The addition of the Polyakov loop is therefore necessary in 
order to quantitatively obtain the required rise in the pressure towards the 
Stefan--Boltzmann limit of QCD corresponding to the horizontal dashed line in the figure. 
The difference between the chiral model pressure and the additional increase due to the 
Polyakov loop can be understood by looking at the contribution of bosons and fermions in 
the ideal gas result which for zero chemical potentials is 
\begin{displaymath}
p_{\rm{SB}}=\frac{\pi^2 T^4}{45}\left((N_c^2-1) +\frac{7N_c N_f}{4}\right).
\end{displaymath}
Setting $N_c=3$ and $N_f=2$, the ratio of the bosonic and fermionic contributions is 
$g_B/g_F=16/21\approx 0.76$. 
 
Both PNJL and PLSM models give a good overall desctiption of QCD pressure above $T_c$. Let 
us then turn to the analysis of more differential observables. The entropy density 
$s(T)=p^\prime(T)$, the energy density $\epsilon(T)=Ts(T)-p(T)$, trace of the energy 
momentum tensor $\theta^\mu_{~\mu}(T)=\epsilon(T)-3p(T)$ and the heat capacity 
$c(T)=\epsilon^\prime(T)=Tp^{\prime\prime}(T)$, to name a few, serve as important probes 
which, together with lattice data, allow to probe the validity of different models and the 
knowledge of the temperature dependence of these quantites is important for phenomenology 
of relativistic heavy ion collisions as well as for cosmology. 

Let us first consider the trace of the energy momentum tensor. We show the result of 
different models in right panel of Fig. \ref{pressure_traceanomaly}. Notice how also here 
the inclusion of the Polyakov loop is imperative to match lattice data in comparison to LSM 
and NJL models: First, the addition of the Polyakov loop removes the qualitatively 
different structures in LSM and NJL model around $T_c$ and in the coupled models a smooth 
universal curve results. Second, the inclusion of the Polyakov loop is important also for 
obtaining agreement with the asymptotic behavior above $T_c$. In fact both LSM and NJL 
models alone predict that at high temperatures $\theta^\mu_{~\mu}(T)\times 
T^4={\rm{const.}}$ contrary to the observation. 

\begin{figure}[htb]
\centering
\includegraphics[width=8.5cm]{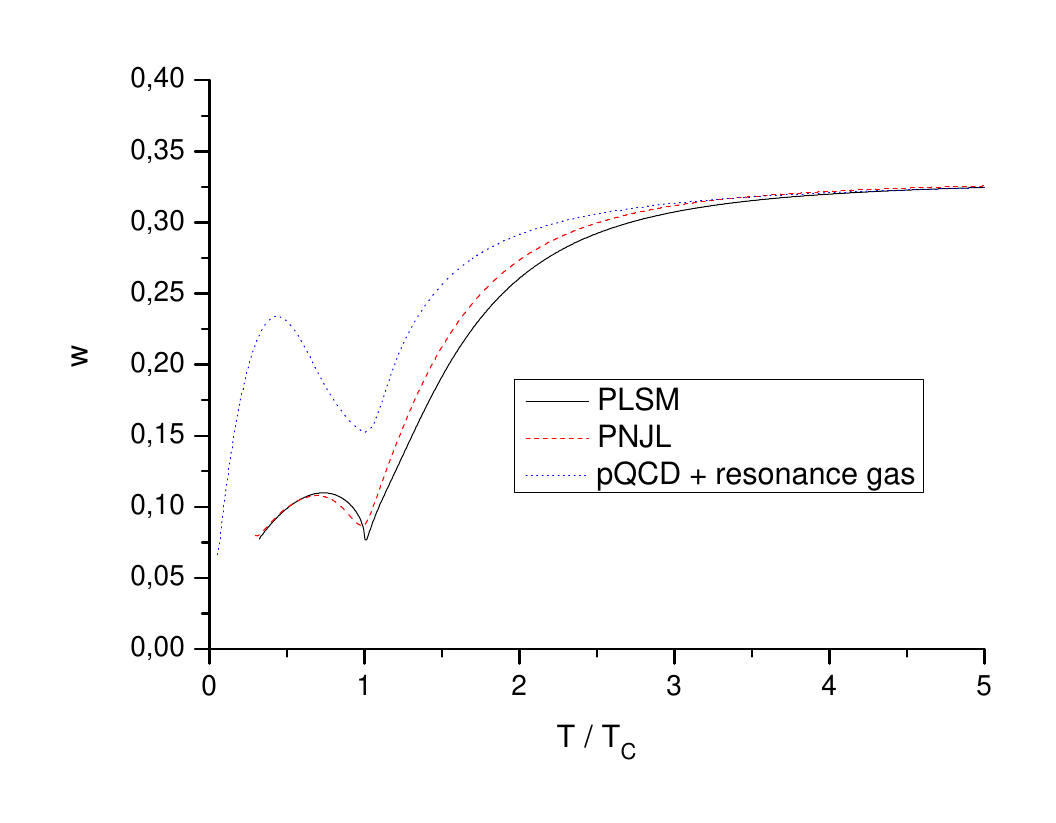}
\caption{The equation of state parameter $w(T)=p(T)/\epsilon(T)$.}
\label{eos_parameter}
\end{figure}

The information contained in the trace anomaly can also be represented in terms of the 
"equation-of-state -parameter" $w(T)=p(T)/\epsilon(T)$ which we show in Fig. 
\ref{eos_parameter}. Both PLSM and PNJL models give again very similar results, 
significantly lower than the pQCD result also shown in the figure. The effective model 
results are similar to recent lattice data \cite{Ejiri:2005uv}. 
There are two clear features deserving further numerical and theoretical studies: First, 
the drop near $T_c$ leads to $w\sim 0.1$ in PNJL and LSM models as well as in the lattice 
data while the perturbation theory result is larger, $w\sim 0.15$. Second, below $T_c$ the 
pQCD model leads to larger rise in $w(T)$ than the lattice data and PNJL and PLSM models. 
These features are likely to be affected by the resonance gas dynamics neglected in the 
effective models but which are present in the pQCD model. 

\begin{figure}[htb]
\centering\hspace{-0.3cm}
 \subfigure{
  \includegraphics[width=5.7cm]{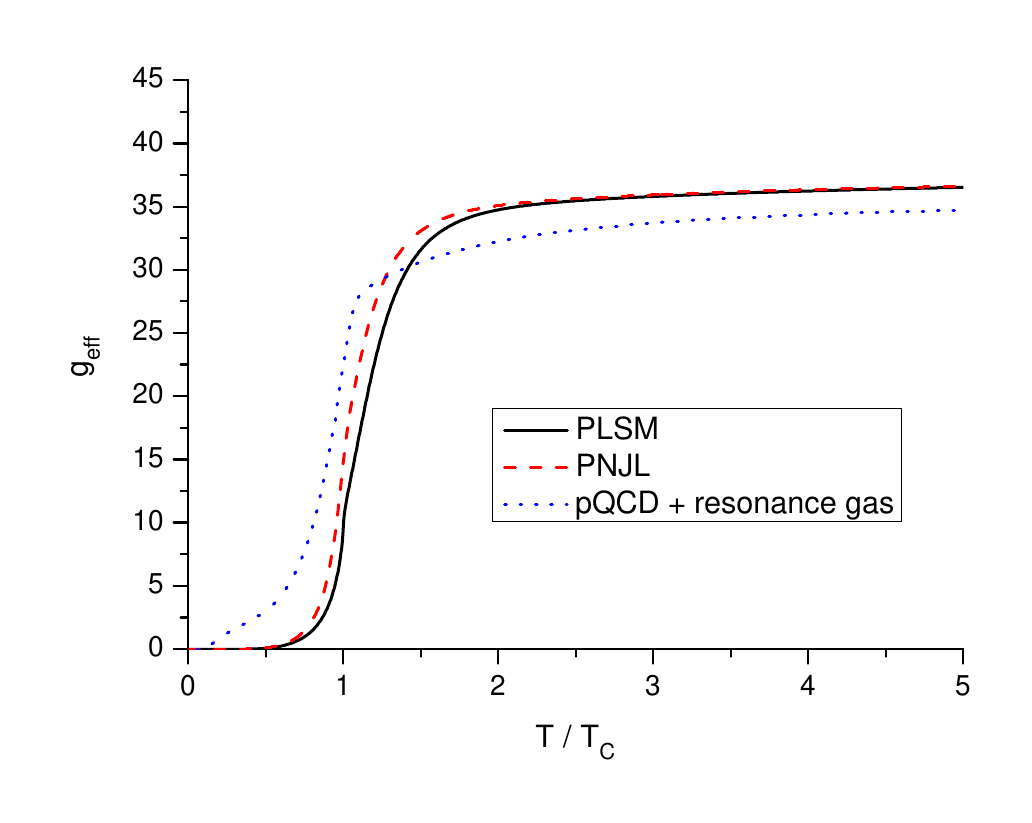}}
	\subfigure{\includegraphics[width=5.7cm]{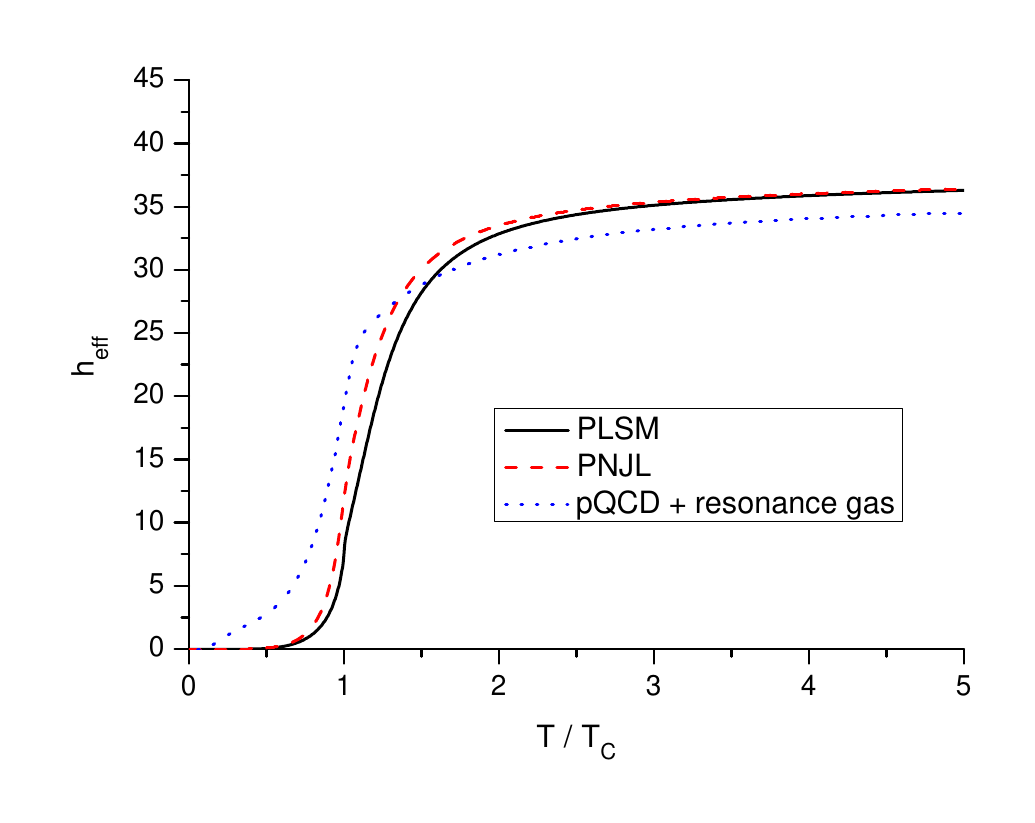}}
  \subfigure{\includegraphics[width=5.7cm]{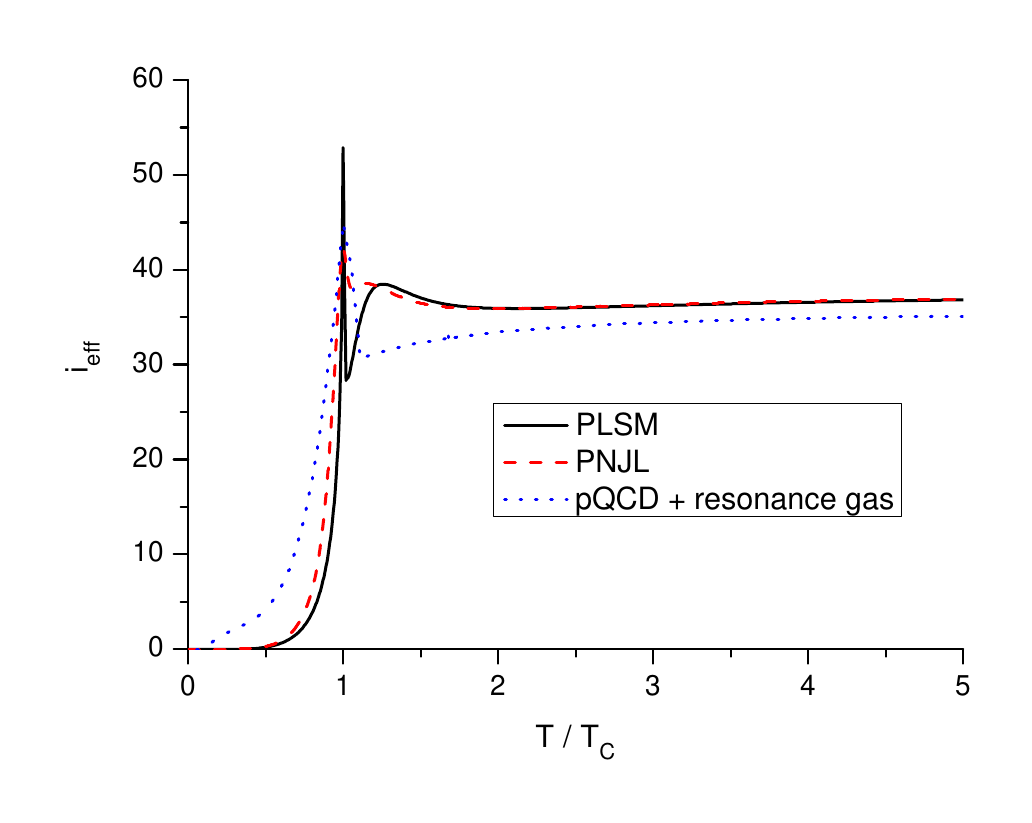}}
\vskip-0.4truecm
\caption{{\it Left Panel}: $g_{\rm{eff}}(T)$ at $\mu=0$. {\it Middle Panel:} 
$h_{\rm{eff}}(T)$ at $\mu=0$. {\it Right Panel:}  $i_{\rm{eff}}(T)$ at $\mu=0$.}
\label{effdofs}
\end{figure}

The information contained in the derivatives defined above can be presented and explored in 
various ways. We choose to follow the presentation in \cite{Laine:2006cp}, since that 
allows also a quantitative comparison against the results from the resummed perturbation 
theory. In Fig. \ref{effdofs} we plot the effective numbers of degrees of freedom defined 
by
\begin{eqnarray}
g_{\rm{eff}}\equiv\frac{\epsilon(T)}{\left[\frac{\pi^2T^4}{30}\right]},~~h_{\rm{eff}}\equiv
\frac{s(T)}{\left[\frac{2\pi^2T^3}{45}\right]}, 
~~i_{\rm{eff}}\equiv\frac{c(T)}{\left[\frac{2\pi^2T^3}{15}\right]}
\end{eqnarray}

Looking at the effective degrees of freedom, we again see that both PLSM and PNJL model 
results are consistent with each other and with the corresponding QCD+resonance gas 
results. Only $i_{\rm{eff}}$ shows some qualitative differences 
in how these two models respond to finite temperature as the peak at $T_c$ is sharper in 
PLSM model than in PNJL model. This peak arises since $i_{\rm{eff}}$ is 
proportional to heat capacity which diverges in a second order phase transition. Another 
issue present in the temperature dependence of $i_{\rm{eff}}$ is a small second peak 
visible very weakly in PNJL model but more strongly in PLSM model. This is due to the 
remnant of the deconfinement transition described by the temperature dependence of the 
parameters in the Polyakov loop potential. Note how it is vital to look at more 
differential observables, second derivatives in this case, to see this effect. More precise 
lattice data is needed to determine if the structure of two independent phase transitions 
connected by interactions underlying these effective models is indeed correct.

\subsection{Thermodynamics at $\mu\neq 0$: location of the critical point?}

Let us then consider the consequences of nonzero net quark density by allowing for finite 
chemical potential. As a starting point we will use the grand canonical potential derived 
in section \ref{section_models}, and neglect possible direct $\mu$-dependence of the 
Polyakov loop potential which has been discussed eg. in ref. \cite{Schaefer:2007pw}. 

\begin{figure}[htb]
\centering
  \subfigure{
  \includegraphics[width=7cm]{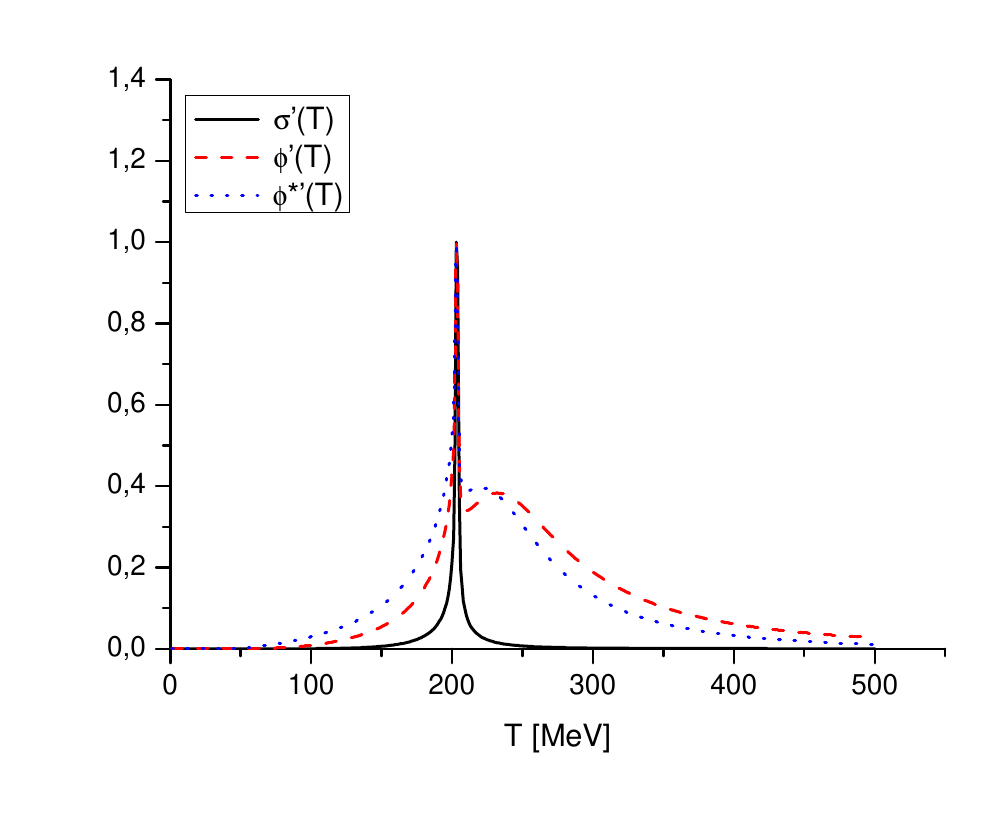}}
	\qquad
	\subfigure{\includegraphics[width=7cm]{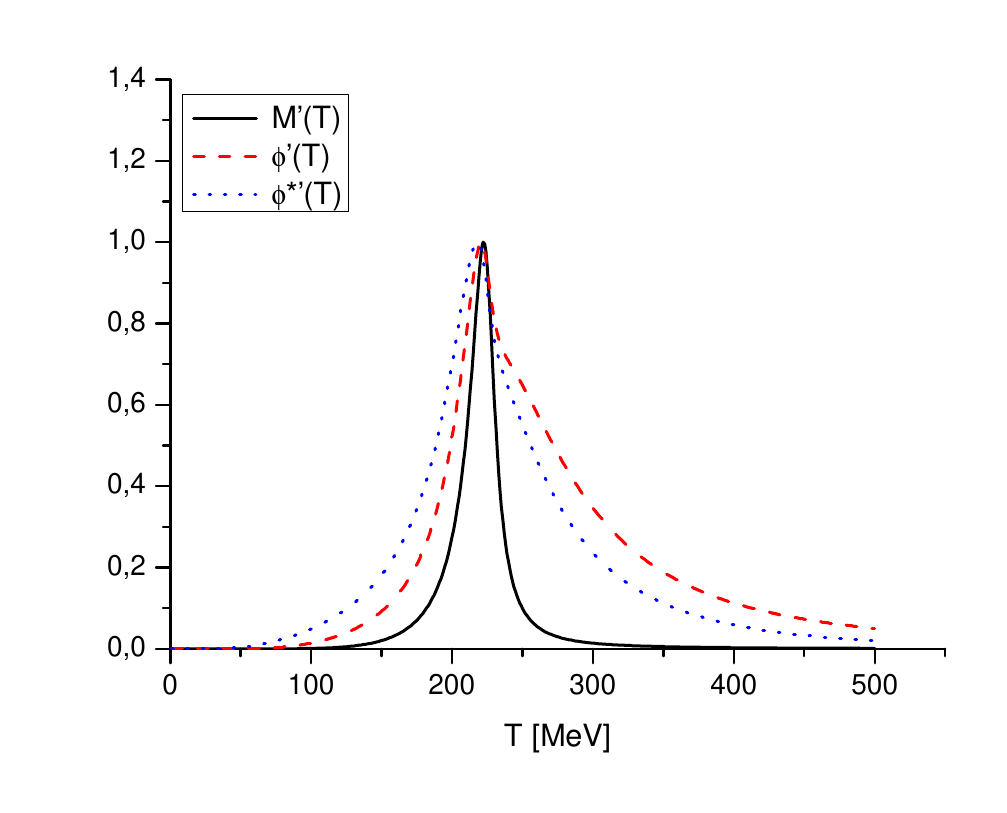}}
\vskip-0.4truecm
\caption{{\it Left Panel}: Temperature derivatives of the mean fields $\phi(T)$ and 
$\sigma(T)$ at $\mu=100$ MeV in the PLSM model.  {\it Right Panel:} Same observables as in 
the left panel but in the PNJL model.}
\label{condensates_mu100}
\end{figure}

\begin{figure}[htb]
\centering
  \subfigure{
  \includegraphics[width=7cm]{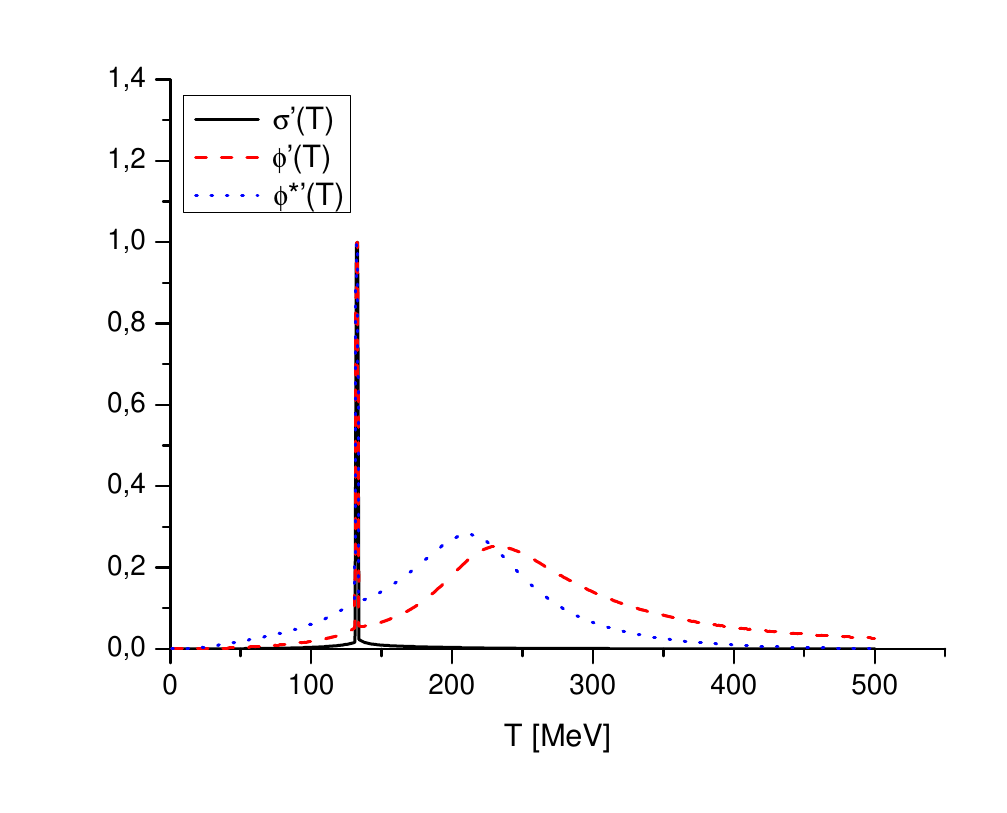}}
	\qquad
	\subfigure{\includegraphics[width=7cm]{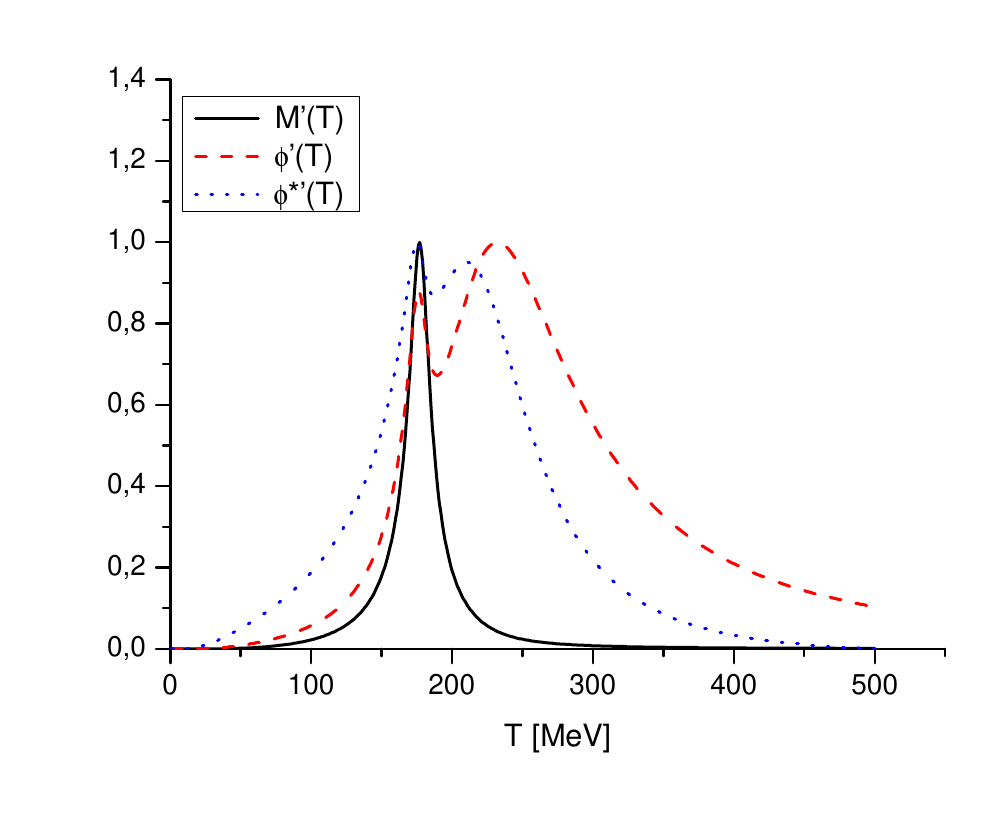}}
\vskip-0.4truecm
\caption{{\it Left Panel}: Temperature derivatives of the mean fields $\phi(T)$ and 
$\sigma(T)$ at $\mu=250$ MeV in the PLSM model.  {\it Right Panel:} Same observables as in 
the left panel but in the PNJL model.}
\label{condensates_mu250}
\end{figure}

First we study if the coincidence of deconfinement and chiral symmetry restoration holds 
also at finite chemical potential. In Fig. \ref{condensates_mu100} we show the derivatives 
of the condensates as a function of temperature at $\mu=100$ MeV and in Fig. 
\ref{condensates_mu250} at $\mu=250$ MeV. 
We observe a coincidence of peaks in the derivatives of the condensates in both of these 
effective models. Note that at finite chemical potential $\phi$ and $\phi^\ast$ are no more 
equal.
At large chemical potentials we observe that already at first derivatives of the 
condensates a double peak structure arises. This is due to the fact that as chemical 
potential is increased, the critical temperature in the chiral sector decreases as shown by 
the location of the leftmost peak in the figures, while the remnant of the deconfinement 
transition in the Polyakov loop potential is unaffected by the value of the chemical 
potential and remains visible at temperature $T\sim 200$ MeV. As the two distinct 
transitions underlying these effective models are separated over wider temperature range 
their separate features become also more visible. Hence, this provides another way to 
numerically investigate the correctness of the initial assumption of independent 
deconfinement and chiral symmetry restoration underlying these models. 

\begin{figure}[htb]
\centering

\includegraphics[width=10cm]{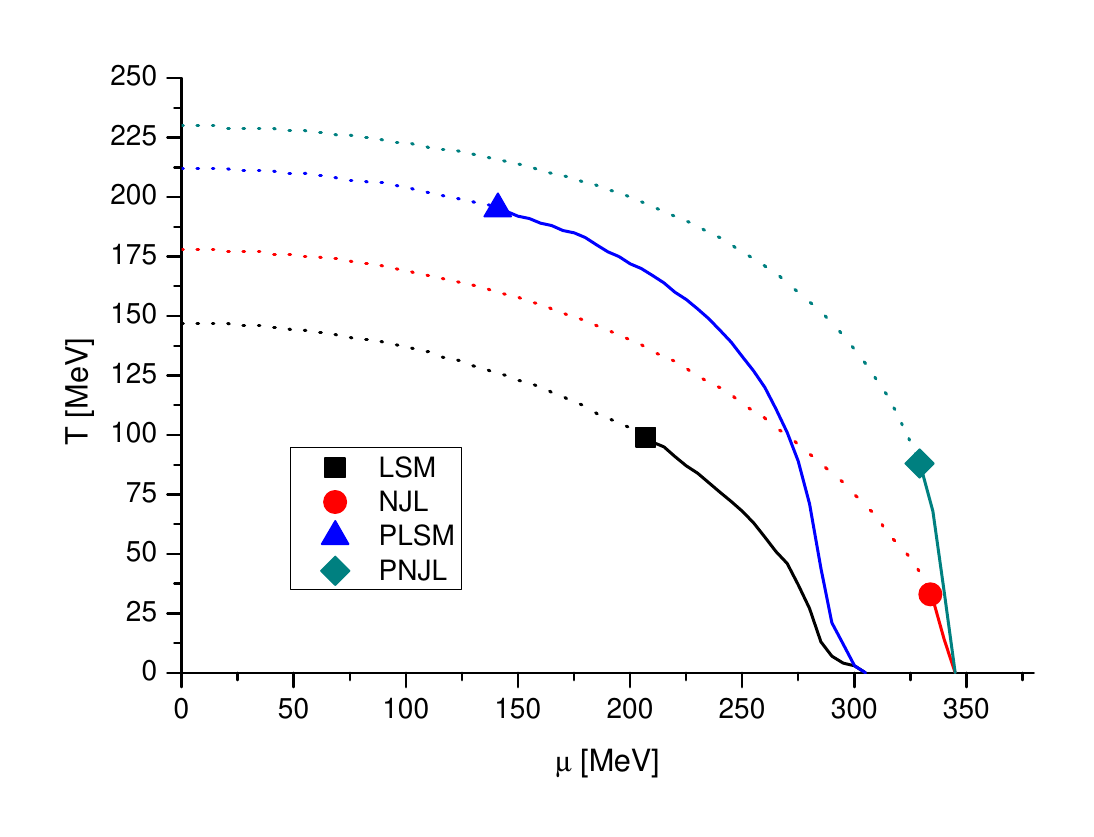}
  
\vskip-0.4truecm
\caption{The $(T,\mu)$ phase diagram.  Leftmost pair of curves shows the result for LSM and 
PLSM while the second pair is for NJL and PNJL models. The solid part of the curve denotes 
a first order transition while the dashed part is a crossover.}
\label{phase_diagram}
\end{figure}

We show the $(T,\mu)$ phase diagram in Fig. \ref{phase_diagram}, where the solid line shows 
the result of PLSM and the dotted line shows the result of PNJL model. The lines have been 
determined by finding the location of the peak in the chiral condensate $\sigma(T)$ at each 
chemical potential. Low temperature and low net quark density phase is confined and chiral 
symmetry is broken, while at high temperature the chiral symmetry is restored and the 
system is deconfined. The two models yield very different values for the location of the 
critical point: while the PLSM gives $(T_c,\mu_c)=(195,141)$ for the position of the 
critical point, the PNJL model yields (88,329). This can be due to several reasons, e.g. 
the neglect of the possible chemical potential dependence of the Polyakov loop potential. 
On the other hand, this can be due to neglect of possible relevant degrees of freedom. It 
is likely that at finite chemical potential diquark degrees of freedom become important and 
should be taken into account; see  \cite{Roessner:2006xn}. This situation would be similar 
to the differences between LSM and NJL models at finite temperature which were shown to 
reduce once the Polyakov loop dynamics is accounted for. 

To discover the critical point experimentally in heavy ion collisions, it would be 
desirable to have a reliable quantitative theoretical estimate of its location. As we have 
explicitly seen here, various effective theory estimates deviate a lot when finite chemical 
potentials are considered. The lattice determinations using different techniques also lead 
to very different results for the location of the critical point. For example, for two 
flavors the authors of \cite{Fodor:2004nz} find the critical point at $\mu_B\sim 360$ MeV 
while in \cite{Gavai:2004sd} a value $\mu_B\sim 180$ MeV is reported. Currently the 
existence of the critical point is debatable \cite{Philipsen:2007rm,deForcrand:2007rq}, and 
in any case one should be careful in drawing any conclusions from $N_f=2$ results to the 
physical 2+1 case . 

\begin{figure*}[htb]
\centering
  \subfigure{
  \includegraphics[width=7.6cm]{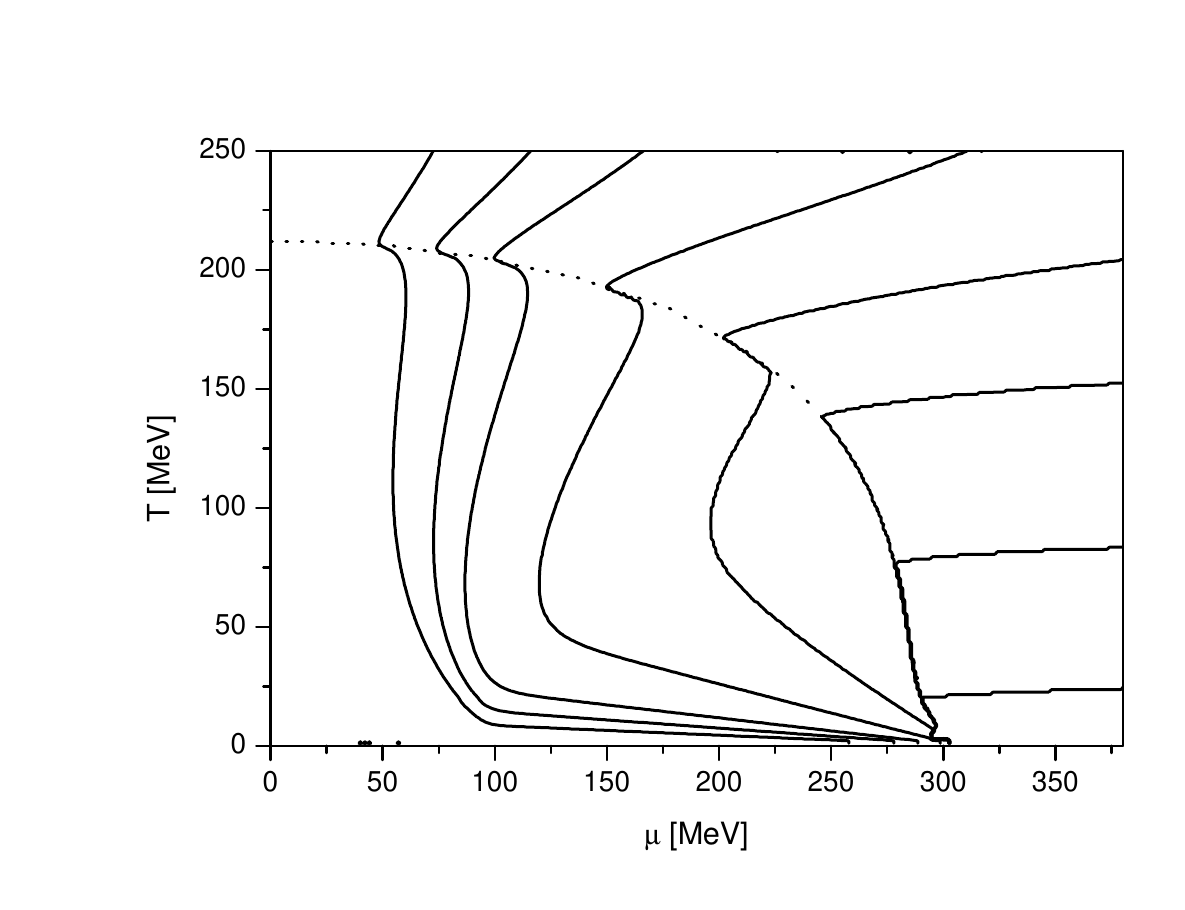}}
  \qquad
  \subfigure{\includegraphics[width=7.4cm]{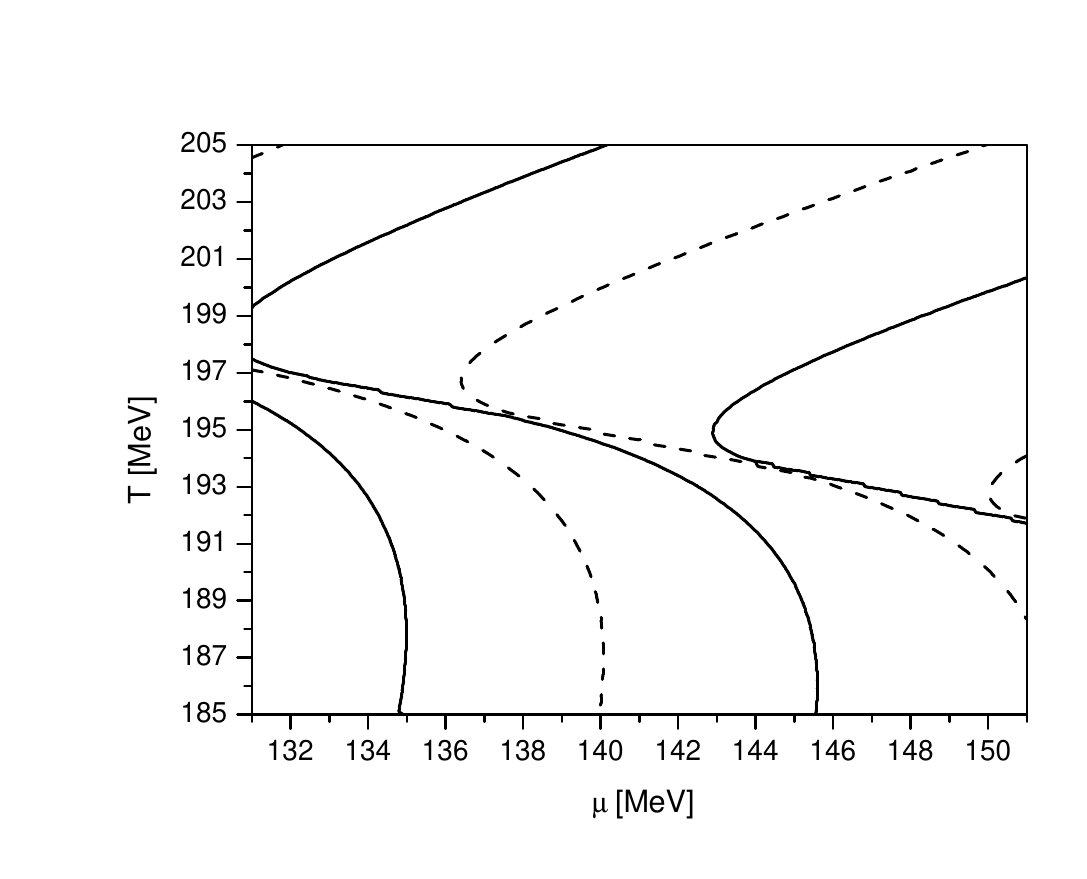}}
\vskip-0.4truecm
\caption{{\it Left Panel}: Constant $S/A$ curves in PLSM model. {\it Right Panel:} A 
closeup on the critical point at $(T_c,\mu_c)=(195,141)$. Every second curve has been drawn 
with the dashed line only to enhance readability.}
\label{isentropes_PLSM}
\end{figure*}

\begin{figure*}[htb]
\centering
  \subfigure{
  \includegraphics[width=7.6cm]{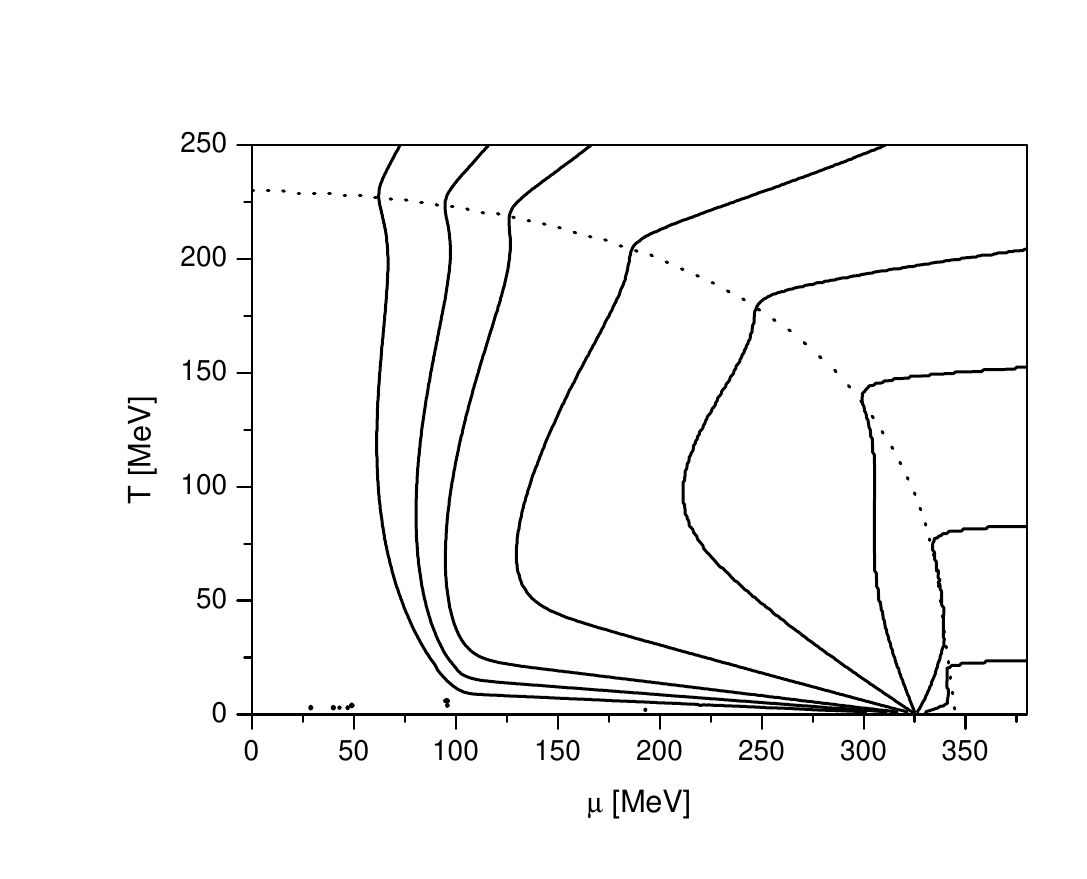}}
  \qquad
  \subfigure{\includegraphics[width=7.4cm]{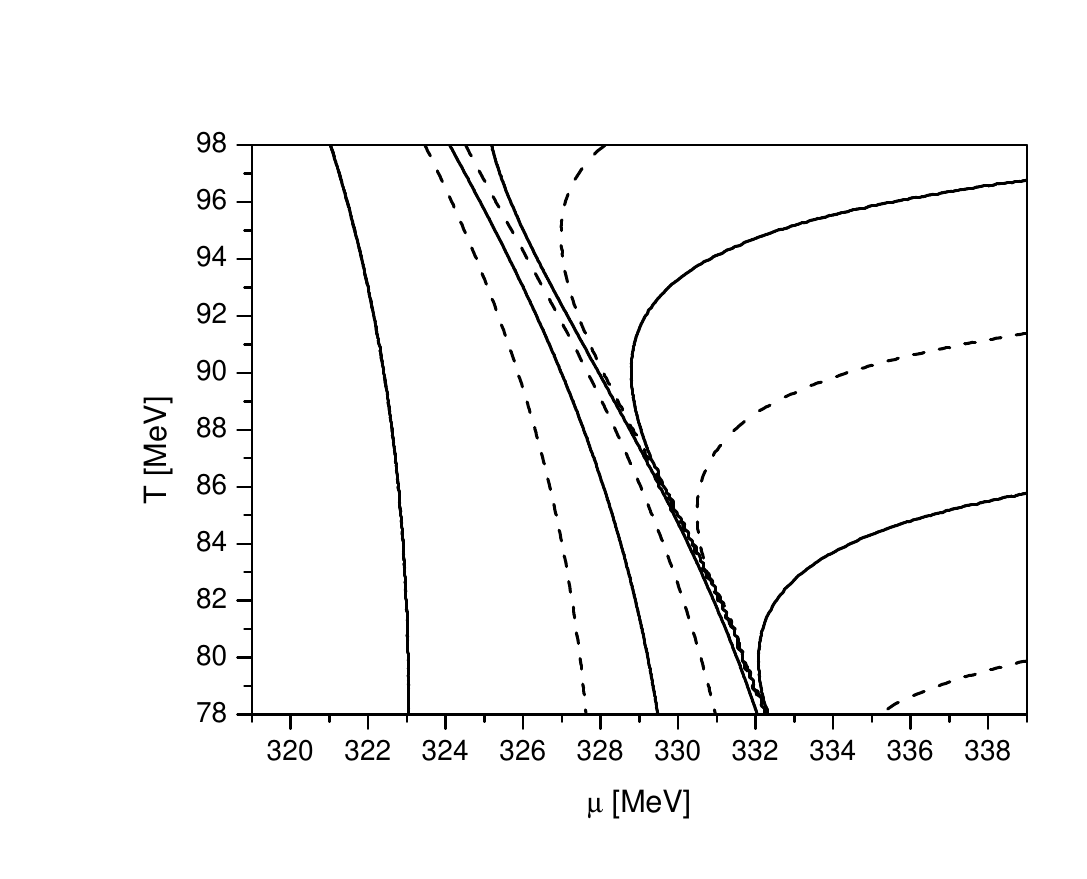}}
\vskip-0.4truecm
\caption{{\it Left Panel}: Constant $S/A$ curves in PNJL model. {\it Right Panel:} A
closeup on the critical point at $(T_c,\mu_c)=(88,329)$.}
\label{isentropes_PNJL}
\end{figure*}

With these remarks in mind, let us assume that the critical point in the $(T,\mu)$ plane 
exists as implied by these effective theories. Then, even if the exact location of the 
critical point is not exactly known, one may argue in favor of its experimental detection 
if the spacetime evolution of the strongly interacting elementary particle matter is such 
that the system passes through the vicinity of the critical point starting from almost any 
initial condition. The outcomes of such {\em{focusing}} behavior have been recently 
advocated for in \cite{Lacey:2006bc} strongly motivated by \cite{Nonaka:2004pg}. However, 
the focusing observed in \cite{Nonaka:2004pg} can be due to the particular equation of 
state applied in that work and is not a general feature of hydrodynamics approach applied 
successfully to describe the RHIC data as in e.g. \cite{Eskola:2005ue}. Given the success 
of ideal fluid hydrodynamics in the description of the RHIC data, it is likely that the 
system expands nearly isentropically. Therefore, to decide if the focusing behavior in the 
models studied here should occur, we find the adiabats of PLSM and PNJL models in the 
$(T,\mu)$ plane. The result is shown in Figs \ref{isentropes_PLSM} and 
\ref{isentropes_PNJL}, and the conclusion for both of these models is negative. Based on 
these figures we conclude that for the hydrodynamical evolution starting from a near zero 
net quark density, as would be the case in a central Au+Au collision at RHIC, the 
trajectory of the system in $(T,\mu)$-plane is non-focusing. As the closeups in the right 
panels of Figs \ref{isentropes_PLSM} and \ref{isentropes_PNJL} show, there is no special 
behavior near the critical points. This result is similar to the one obtained on the 
lattice \cite{Ejiri:2005uv}.     

\begin{figure*}[htb]
\centering
  \subfigure{
  \includegraphics[width=7.6cm]{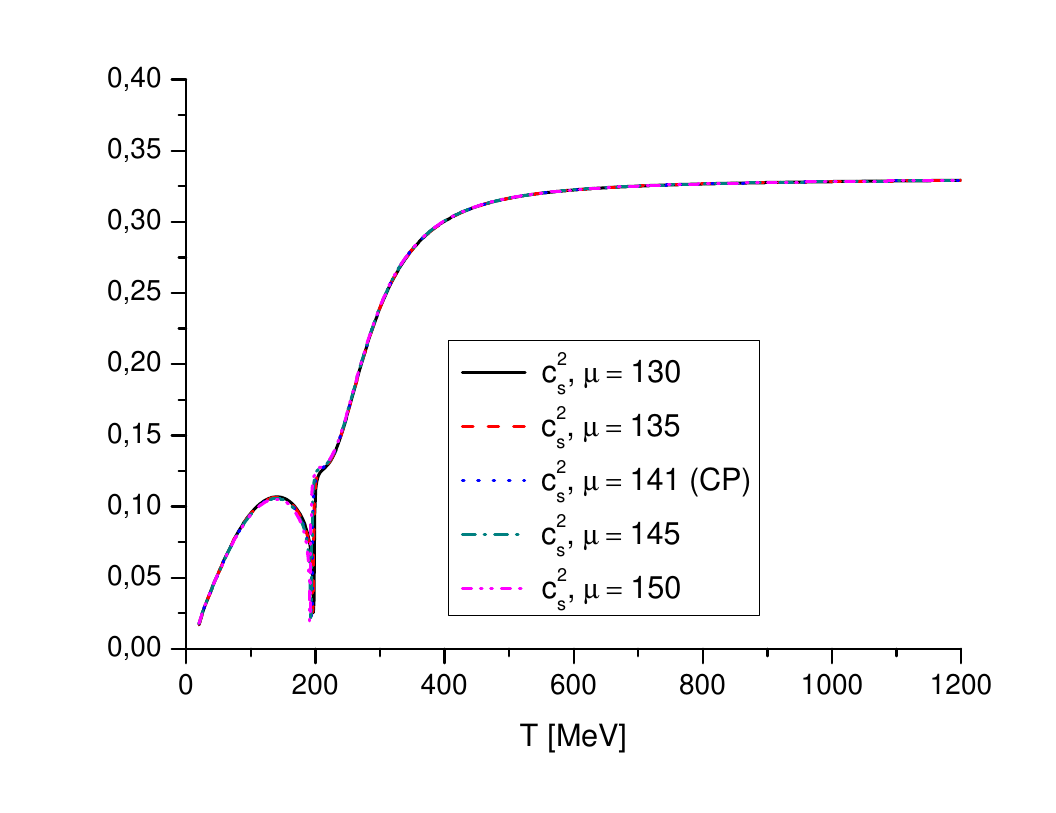}}
  \qquad
  \subfigure{\includegraphics[width=7.4cm]{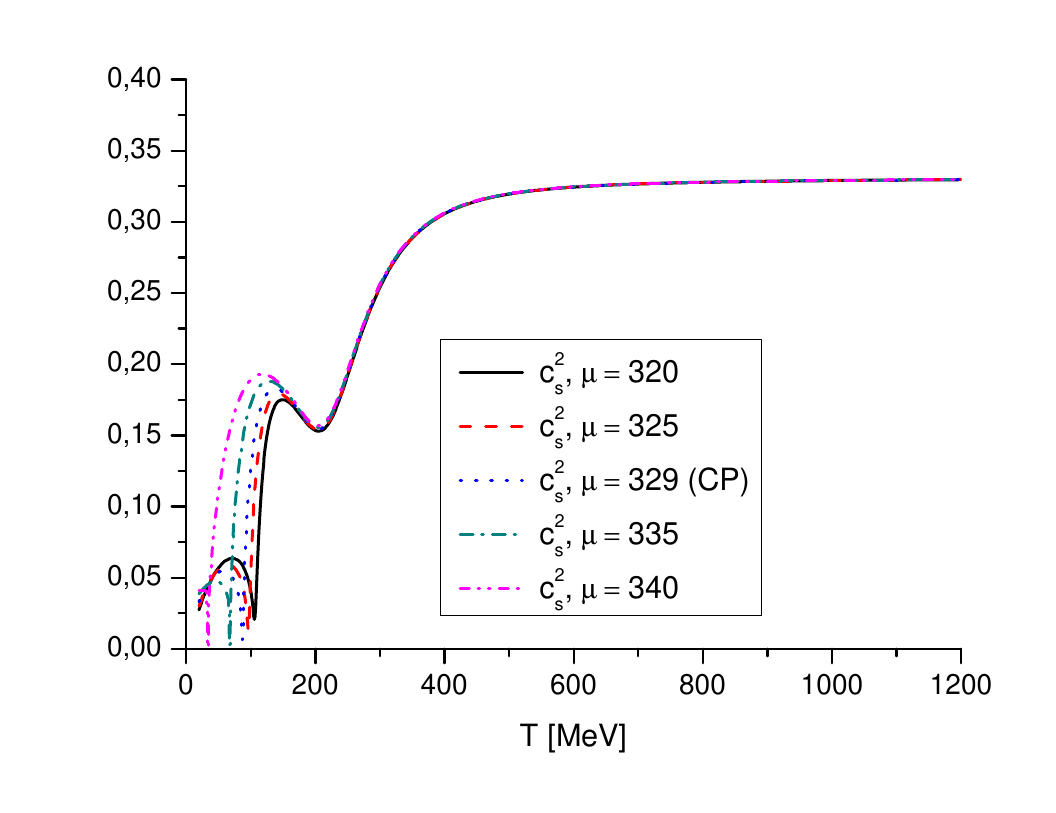}}
\vskip-0.4truecm
\caption{{\it Left Panel}: Sound speed at different chemical potentials in PLSM model. {\it 
Right Panel:} Same for the PNJL model.}
\label{sound_speed}
\end{figure*}

Another interesting quantity is the sound speed shown in Fig. \ref{sound_speed} for 
chemical potentials near the critical one. The stronger dependence of the location of the 
minimum in $c_s(T)$ as a function of the chemical potential in PNJL model is due to the 
fact that near the critical point
the value of $T_c(\mu)$ changes more rapidly than in the case of PLSM model. Together with 
Figs. \ref{isentropes_PLSM} and \ref{isentropes_PNJL} we see from here that in the 
hydrodynamical evolution the sound speed is very small during the part of the evolution the 
system spends near the phase transition region. At $(T,\mu)$ regions relevant for RHIC and 
LHC/ALICE phenomenology, the sound speed is probably well approximated by the $\mu=0$ 
result as the system evolves very close to the $\mu\sim 0$ axis along a nearly parallel 
trajectory until low temperatures deep in the hadronic phase are reached and the trajectory 
bends to end at the finite value $\mu_{\rm{vac}}$ at zero temperature. However, if it was 
possible to create systems which would follow the trajectory bending along the phase 
boundary as the ones at large chemical potential in Figs \ref{isentropes_PLSM} and 
\ref{isentropes_PNJL}, then the sound speed could stay small over larger temperature range 
and this might have consequences for e.g. Mach cones created by high momentum jets 
traversing the thermal medium \cite{Renk:2005ta}. 
On the other hand, while the collisions planned at GSI/FAIR facility might lead to more 
optimal $S/N$ trajectories for this argument to work, the probability for the production of 
the required high momentum probes is much smaller. However, we remind the reader that a 
thorough investigation of the heavy ion phenomenology should be carried out with an 
effective theory suited to describe the $N_f=2+1$ case.

\section{Conclusions and outlook}
\label{section_checkout}

In this paper we have analysed thermodynamics of two effective theories for QCD containing 
as relevant degrees of freedom the Polyakov loop and chiral fields within a framework 
proposed to correctly interpolate between the pure gauge, center symmetric, and chirally 
symmetric two flavor QCD. At zero chemical potential but finite temperature we observed 
that the contribution from the Polyakov loop to thermodynamical quantities like the 
pressure is very important; roughly half of the total pressure. Another aspect which 
underlines the importance of the Polyakov loop sector is the fact that adding it to the 
dynamics tends to diminish the qualitative differences present in the bare chiral theories 
at finite temperature. We have considered explicitly two realizations, the linear sigma 
model and the NJL model. Our results indiate that even if both LSM and NJL models describe 
the chiral dynamics of the QCD vacuum correctly, they must be supplied by other degrees of 
freedom in order to obtain quantitatively correct effective description of QCD 
thermodynamics. We presented a comparison with respect to a recent result interpolating 
between resummed perturbation theory result and resonance gas result \cite{Laine:2006cp}, 
and outlined how future lattice simulations could allow one to obtain more insight into the 
QCD dynamics near the phase transition.

We also considered finite chemical potentials and determined the ($\mu,T$)-phase diagram of 
these theories. Here we observed large discrepancies between the PLSM and PNJL models. 
Based on the above discussion, an obvious explanation would be that some important dynamics 
is again being missed. A natural candidate for a new relevant degree of freedom is the 
diquark condensate responsible for the color superconducting phenomena. We aim to extend 
our work towards this direction next. We also determined the lines of constant $S/A$ in 
$(T,\mu)$-plane and discussed the implications for the equation of state as well as for the 
possible focusing behavior relevant for the experimental discovery of the QCD critical 
point. 

These results can be used for phenomenological applications. The parametrization of the 
equation of state obtained from these models has been shown to agree with lattice data at 
$\mu=0$ but is easily evaluated also at finite values of the chemical potential. Hence it 
could be applied in hydrodynamical simulations of ultrarelativistic heavy ion collisions. 
We have evaluated the sound speed, and shown that on the hydrodynamically relevant 
trajectories there may be substantial temperature range over which the sound speed is 
small. However, it may prove difficult in the laboratory to create systems which would 
follow these particular trajectories. We hope to extend these phenomenological studies 
within realistic hydrodynamics in near future.

There are several improvements to be addressed. We have already stressed that below the 
critical temperature, a more careful treatment of hadronic degrees of freedom is required, 
and the lack of this treatment is best seen in the apparent underestimate of the pressure 
below $T_c$. Another issue clearly concerns the number of active flavors. Here we have 
concentrated only on the case of two flavors, while for more quantitative phenomenology it 
is vital to have the effects of the strange quark under control. However, as a first 
approximation for the equation of state at small chemical potential it should be reasonable 
to multiply the pressure by the overall factor $g_{\rm{SB}}(N_f=3)/g_{\rm{SB}}(N_f=2)$ 
where $g$ counts both the bosonic and fermionc degrees of freedom.
Lattice data \cite{Karsch:2000ps} indicates that the flavor dependence of the QCD pressure 
is dominated by the Stefan-Boltzman factor and the above scaling can be applied within 10 
\% accuracy. Similar reasoning has been used also in \cite{Laine:2006cp} to obtain the 
perturbative QCD result for full $N_f$ flavors with masses $\{m_{N_f}\}$.  We aim to 
address these issues in future work within the effective theory framework discussed here.

\acknowledgements
We thank A.Dumitru and T. Renk for useful discussions and K.J.Eskola for careful reading of 
the manuscript. T.K. thanks the Vaisala foundation for financial support.



\end{document}